\documentstyle[psfig,11pt,newpasp,twoside]{article}
\index{summary}
\index{instructions}
\index{template}

\def\edcomment#1{\iffalse\marginpar{\raggedright\sl#1\/}\else\relax\fi}
\reversemarginpar

\begin{document}
\title{High-Energy Gamma Rays from Neutron Stars in Supernova Remnants: 
From EGRET to GLAST
}
 \author{D.J. Thompson}
\affil{Laboratory for High Energy Astrophysics, NASA Goddard Space Flight Center, Greenbelt, MD 20771}
\author{S.W. Digel}
\affil{Laboratory for High Energy Astrophysics/USRA, NASA Goddard Space Flight Center, Greenbelt, MD 20771}
\author{P.L. Nolan}
\affil{Hansen Experimental Physics Laboratory, Stanford University, Stanford, CA 94305}
\author{O. Reimer}
\affil{Laboratory for High Energy Astrophysics/NRC, NASA Goddard Space Flight Center, Greenbelt, MD 20771}

\begin{abstract}
At least three pulsars in supernova remnants were detected at E $>$ 100 MeV by EGRET on the Compton Gamma Ray Observatory.  Efforts to search for additional pulsars in the EGRET data have been unsuccessful due to limited statistics.  An example is the recently-discovered radio pulsar J2229+6114, where efforts to search the EGRET data using several different methods failed to find significant evidence of pulsation. The GLAST Large Area Telescope (LAT) will have a much greater effective area and a narrower point-spread function than EGRET.  In addition, the field of view will be more than 4 times larger than EGRET's, and the LAT will scan to avoid occultation by the earth, increasing by a large factor the total number of photons detected. The greater rates of photons from pulsar candidates and better discrimination of diffuse interstellar emission will enhance the sensitivity of pulsation searches.   These improvements also offer the prospect of resolving point sources from extended emission in some SNR to define the nature of the associations of EGRET sources with SNR.  Further, work with the GLAST LAT will benefit from ongoing multiwavelength studies (e.g., for RX 1836.2+5925) that provide specific candidate targets for gamma-ray studies. 
\end{abstract}

\section{Introduction - EGRET Pulsars in Supernova Remnants}

At least three of the rotation-powered pulsars seen in high-energy gamma rays by EGRET (Crab, Vela, PSR B1951+32) are neutron stars in SNR.  A fourth, PSR B1706$-$44, would be expected from its age to have a visible SNR, but the observational evidence is unclear.  (Dodson, Gvamaradze, these proceedings.) Because EGRET and previous gamma-ray telescopes lacked the resolution to distinguish neutron star emission from nebular emission, the only certain detections are based on pulsed emission.  

\begin{figure}
\centerline{\psfig{figure=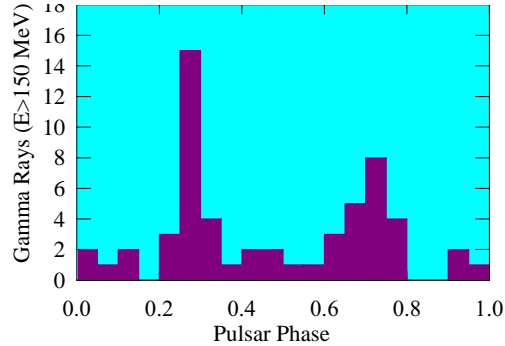,height=4.5cm,bbllx=15pt,bblly=15pt,bburx=570pt,bbury=370pt,clip=.}}
\caption{Possible EGRET light curve for PSR J2229+6114 in Viewing Period 34, based on a search of periods and period derivatives consistent with the later radio measurements.}
\end{figure}

\section{EGRET Search for PSR J2229+6114}

The recent discovery by Halpern et al. (2001) of PSR J2229+6114, a young, high spindown rate radio and X-ray pulsar in an EGRET source error box strongly suggests an association with the EGRET source based on energetics (Halpern, these proceedings).  We have applied three approaches to searching for pulsations in the EGRET data.  Such a search is necessary, because the EGRET observations occurred years before the pulsar discovery, and pulsars in this age range are typically noisy and/or subject to glitches. 
The methods:

1.  We  applied the program described in Jones (1998) and Jones et al. (1997) to the data in VP34, the viewing period with the greatest detection significance. This program uses the ideas of Gregory \& Loredo (1992), extended to take into account the angular distribution of the detected photons and the diffuse backgrounds observed by EGRET.  It calculates a likelihood for each assumed pulsar period, using a 5-bin light curve, marginalizing over the arbitrary choice of zero phase. 

2. We carried out a frequency scan on barycentered event files (100$-$300, 300$-$1000, 100$-$1000, $\ge$1000 MeV) for 10 viewing periods.  H-test and Protheroe-test statistics were applied.

3. We made a pulse folding search on barycentered event files E$>$50 MeV using individual viewing periods, then attempted to link the best results into a coherent timing solution.  The statistics used were the H-test and Rayleigh test. 

The most promising solution is shown in Fig. 1, for VP34.  The energy and angle cuts have been optimized to maximize the significance.  The resemblance of this light curve to others seen in EGRET data (two peaks separated by 0.4$-$0.5 phase) is intriguing, but the other viewing periods, all of which had lower statistics but were closer in time to the observed radio and X-ray data, produced no consistent, confirming signal.  By itself, even this optimized  signal is of marginal significance ($\ge$1\% probability of chance occurrence) based on the trials factor. We do {\bf not} claim a detection.

\section{Searching for Pulsars with GLAST}

The limitation on searches of the EGRET data is statistical:  the number of source photons in excess of the diffuse gamma radiation is too small.  All three of the techniques described above have demonstrated the ability to find pulsations for sufficiently bright sources. The GLAST Large Area Telescope (LAT) will have a much greater effective area (factor of $\sim$6 at 1 GeV) and a narrower point-spread function (factor of $\sim$3 at 1 GeV) than EGRET.  In addition, the field of view will be more than 4 times larger than EGRET's, and the LAT will scan to avoid occultation by the earth, increasing by a large factor the total number of photons detected. The greater rates of photons from pulsar candidates and better discrimination of diffuse interstellar emission will enhance the sensitivity of pulsation searches.    Analyses of LAT performance (e.g., Chandler et al. 2001; Carrami\~{n}ana 2001) show that periodicities should be detectable if present in any of the known low-latitude EGRET sources. 

\begin{figure}
\centerline{\psfig{figure=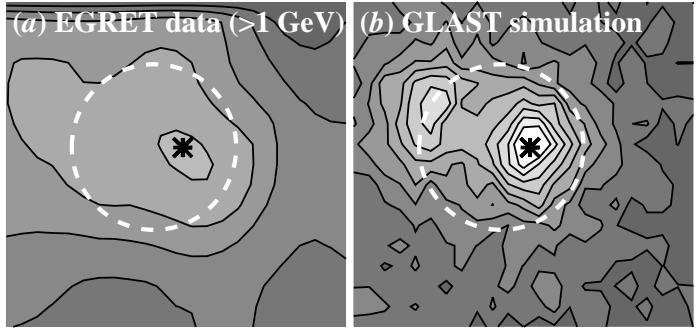,height=4.5cm,bbllx=2pt,bblly=2pt,bburx=330pt,bbury=160pt,clip=.}}
\caption{Comparison of EGRET observations with simulated GLAST LAT sky survey data for $\gamma$-Cygni SNR, assuming a pulsar and nebular component (see text). The dashed circle indicates the extent of the SNR shell and the asterisk the location of a suspected pulsar.}
\end{figure}

\section{Imaging SNR with GLAST}
Observations with the LAT on GLAST should clarify the nature of the gamma-ray emission of EGRET sources positionally associated with SNR as well as fainter SNR/plerion/pulsar sources.  The limits for resolving or closely localizing sources with the LAT depend on the spectral characteristics of the sources, the intensity of the interstellar emission from the Milky Way, and of course the observing strategy, so general statements about resolving power are difficult to make.  The imaging capability of the LAT for SNR is illustrated in Fig. 2 using a model for $\gamma$-Cygni that is consistent with the EGRET observations (see Allen et al. 1999).  The EGRET data suggest spatially extended high-energy emission, and the model assigns 40\% of the flux of the EGRET source to emission from cosmic-ray interactions with interstellar gas at the edge of the SNR shell.

\section{Multiwavelength Indicators for Gamma-Ray Neutron Stars in SNR }
Successful multiwavelength identification campaigns toward unidentified EGRET sources coincident with SNRs often find point sources as more convincing candidates for the observed gamma-ray emission than other emission mechanisms in an SNR. For bright and steady hard spectrum gamma-ray sources such as $\gamma$-Cygni: RX J2020.4+4026/3EG J2020+4017 (Brazier et al. 1996), CTA1: RX J0007.0+7302 / 3EG J0010+7309 (Brazier et al. 1998), IC443: CXOU J061705.3+222127/3EG J0617+2238 (Olbert et al. 2001; Bocchino \& Bykov 2001; Sturner et al. 2001), and AX J1420.1-6049/GEV J1417-6100 (Roberts et al. 2001; D'Amico et al. 2001) there is growing evidence for isolated neutron star/pulsar components.  In cases of narrowly restricted source locations it becomes obvious that the high-energy emission does not correlate well with the radio/X-ray bright shell/rim features of the remnants. Additionally, these gamma-ray sources are often characterized by spectral cutoffs at GeV energies (Reimer \& Bertsch 2001), which resemble the known high-energy cutoffs from pulsars like Geminga, Vela, and Crab. With the superior source location accuracy and the superior sensitivity of GLAST compared to EGRET, multiwavelength campaigns can be executed with complete identifications in the vicinities of high-energy gamma-ray sources.

\end{document}